\documentclass[11pt, twocolumn]{article}

\usepackage{cite}
\usepackage{amsmath, amssymb}
\usepackage{graphicx}
\usepackage{url}

\usepackage{algorithm}
\usepackage{algorithmic}
\usepackage[margin=1in]{geometry}
\usepackage{authblk}
\usepackage{tikz}
\usepackage{listings}
\usepackage{xcolor}

\lstdefinelanguage{Solidity}{
	keywords={typeof, new, true, false, catch, function, return, null, catch, switch, var, if, in, while, do, else, case, break},
	keywordstyle=\color{blue}\bfseries,
	ndkeywords={class, export, boolean, throw, implements, import, this},
	ndkeywordstyle=\color{darkgray}\bfseries,
	identifierstyle=\color{black},
	sensitive=false,
	comment=[l]{//},
	morecomment=[s]{/*}{*/},
	morestring=[b]',
	morestring=[b]"
}

\title{Zer0n: An AI-Assisted Vulnerability Discovery and Blockchain-Backed Integrity Framework}
\date{\today}
\author[1]{Harshil Parmar}
\author[1]{Pushti Vyas}
\author[1]{Prayers Khristi}
\author[1]{Priyank Panchal}
\affil[1]{Parul University, India}

\begin{document}

\maketitle

% =========================
% Abstract
% =========================
\begin{abstract}
As vulnerability research increasingly adopts generative AI, a critical reliance on opaque model outputs has emerged, creating a "trust gap" in security automation. We address this by introducing Zer0n, a framework that anchors the reasoning capabilities of Large Language Models (LLMs) to the immutable audit trails of blockchain technology. Specifically, we integrate Gemini 2.0 Pro for logic-based vulnerability detection with the Avalanche C-Chain for tamper-evident artifact logging. Unlike fully decentralized solutions that suffer from high latency, Zer0n employs a hybrid architecture: execution remains off-chain for performance, while integrity proofs are finalized on-chain. Our evaluation on a dataset of 500 endpoints reveals that this approach achieves 80\% detection accuracy with only a marginal 22.9\% overhead, effectively demonstrating that decentralized integrity can coexist with high-speed security workflows.
\end{abstract}


% =========================
% Main Sections
% =========================
\section{Introduction}
Modern software systems have reached a level of complexity where manual security auditing is no longer sufficient. Consequently, the industry has pivoted toward automated vulnerability discovery, evolving from simple pattern-matching fuzzers to sophisticated reasoners built on Large Language Models (LLMs) \cite{contractfuzzer,intrudtree}. However, this shift introduces a subtle but dangerous risk: the lack of verifiable provenance. When an AI agent reports a vulnerability—or conversely, declares a system secure—there is often no immutable record of that analysis, leaving the process open to manipulation in adversarial environments \cite{weaponizedai,aitrends}.

While recent work has explored securing data pipelines with blockchain \cite{blockchainai,aicyberchain}, these solutions often prioritize data privacy over workflow integrity or introduce prohibitive latency via decentralized storage. We argue that a middle ground is required: a system that leverages the reasoning power of frontier models like Gemini 2.0 Pro while cryptographically binding their outputs to a public ledger.

In this work, we propose Zer0n, an architecture designed to close the trust gap in automated vulnerability research. Rather than offloading the entire analysis to slow on-chain mechanisms, Zer0n maintains a high-speed off-chain analysis engine but enforces integrity through asynchronous blockchain logging. This design avoids reliance on complex decentralized storage systems (e.g., IPFS), preferring a streamlined hashing mechanism that remains practical for real-world latencies. The primary contributions of this article are:
\begin{itemize}
    \item \textbf{A Hybrid Trust Architecture:} We introduce a modular design that decouples AI reasoning from integrity enforcement, allowing each to scale independently.
    \item \textbf{Tamper-Evident Artifacts:} We implement a Solidity-based mechanism on the Avalanche C-Chain that renders post-analysis report manipulation mathematically detectable.
    \item \textbf{Empirical Feasibility Analysis:} We provide a detailed evaluation demonstrating that blockchain integration introduces manageable overhead, making it a viable addition to standard security pipelines.
\end{itemize}

\section{Related Work}

\subsection{AI-Assisted Vulnerability Discovery}
Artificial intelligence and machine learning techniques have been widely applied to intrusion detection and vulnerability analysis \cite{dlsurvey,intrudtree}. Approaches based on decision trees, deep neural networks, and autoencoders have demonstrated effectiveness in detecting anomalous behavior, although challenges related to robustness and interpretability remain \cite{autoencoder}. ContractFuzzer represents an early example of automated vulnerability discovery using fuzzing techniques for smart contracts \cite{contractfuzzer}.

Recent surveys highlight both the potential and the limitations of deep learning for cybersecurity applications, emphasizing issues such as generalization, adversarial manipulation, and operational complexity \cite{dlsurvey}. These observations motivate hybrid approaches that combine automated analysis with complementary mechanisms for trust and verification.

\subsection{Evolution of Smart Contract Analysis}
Traditional smart contract security relies heavily on static analysis and symbolic execution. Tools like \textbf{Slither} \cite{slither} utilize intermediate representations to quickly identify common vulnerability patterns, while \textbf{Mythril} \cite{mythril} employs concolic analysis to explore execution paths for deeper flaws. While effective for known bug classes, these tools often struggle with complex business logic errors that require semantic understanding. Recent comparisons suggest that combining static analysis with AI-driven reasoning can significantly improve detection rates for logic vulnerabilities \cite{contract_comparison}.

\subsection{Generative AI Security Risks}
The rapid adoption of Large Language Models has introduced new security paradigms. The \textbf{OWASP Top 10 for LLM Applications 2025} \cite{owasp_llm} highlights critical risks such as prompt injection, sensitive information disclosure, and system prompt leakage. Furthermore, recent studies indicate that LLMs themselves can be manipulated to produce biased or malicious code analysis results \cite{llm_vuln_2025}. Zer0n addresses this "trust gap" by anchoring LLM-generated analysis artifacts to an immutable blockchain ledger, ensuring provenance and integrity in the face of potential model hallucinations or adversarial compromise \cite{ai_risk_analysis}.

\subsection{Blockchain and Cybersecurity}
Blockchain-based approaches have been proposed to enhance cybersecurity by providing tamper-evident data storage and decentralized trust models \cite{blockchainslr}. Systematic literature reviews indicate that blockchain can improve auditability and accountability but may introduce performance overhead and integration challenges \cite{blockchainsurvey}. Research agendas in blockchain cybersecurity stress the importance of careful system design to avoid unnecessary complexity and misplaced trust assumptions \cite{trustmachine}.

Several works have explored the combination of blockchain and artificial intelligence to secure data pipelines and analytical processes \cite{aicyberchain,blockchainai}. In contrast to domain-specific solutions, Zer0n focuses on the integrity of vulnerability discovery workflows themselves, independent of application domain.

\subsection{Standards and Threat Landscape}
Standardized vulnerability assessment frameworks play a crucial role in operational security. The Common Vulnerability Scoring System (CVSS) remains a widely adopted standard for evaluating vulnerability severity \cite{cvss}. Threat landscape reports published by organizations such as ENISA and guidance from national agencies including BSI highlight the increasing sophistication and scale of cyber attacks, underscoring the need for trustworthy security processes \cite{enisa,bsi}.

\subsection{Comparative Analysis}
Table~\ref{tab:lit_review} contrasts Zer0n against prominent security analysis tools and frameworks discussed in literature. This comparison highlights the unique position of Zer0n effectively bridging the gap between automated reasoning and cryptographic integrity.

\begin{table*}[htbp]
\centering
\caption{Comparison of Zer0n with State-of-the-Art Approaches}
\label{tab:lit_review}
\begin{tabular}{|p{3cm}|p{3cm}|c|c|p{2.5cm}|}
\hline
\textbf{Approach} & \textbf{Core Method} & \textbf{Integrity?} & \textbf{AI?} & \textbf{Focus} \\ \hline
Slither \cite{slither} & Static Analysis (IR) & No & No & Smart Contracts \\ \hline
Mythril \cite{mythril} & Symbolic Execution & No & No & Smart Contracts \\ \hline
AICyberChain \cite{aicyberchain} & Neural Networks & Yes & Yes & IoT Data \\ \hline
\textbf{Zer0n (Ours)} & \textbf{LLM + C-Chain} & \textbf{Yes} & \textbf{Yes} & \textbf{Vuln Logic} \\ \hline
\end{tabular}
\end{table*}

\section{System Architecture}
Zer0n is designed as a modular system composed of three primary components: an AI-assisted vulnerability analysis engine, a blockchain-backed integrity logging layer, and a coordination interface that manages workflow execution and reporting.

The analysis engine is built upon the Google Gemini 2.0 Pro API, leveraging its advanced context window and reasoning capabilities to identify logic flaws and complex vulnerabilities. The workflow follows a five-phase methodology:
\begin{enumerate}
    \item \textbf{Reconnaissance \& Scope:} Intelligent target prioritization and asset mapping.
    \item \textbf{Discovery:} Crawling and technology fingerprinting.
    \item \textbf{Attack Surface Analysis:} Parameter discovery and AI-driven payload generation.
    \item \textbf{Exploitation:} Active validation of potential vulnerabilities using generated PoCs.
    \textbf{Reporting:} Comprehensive aggregation of findings with remediation guidance.
\end{enumerate}
Zer0n utilizes prompt engineering techniques to guide the LLM in simulating sophisticated attack vectors while minimizing false positives. To mitigate model hallucination and ensure consistency, we utilized a temperature setting of 0.2 and structured chain-of-thought system prompts. While the specific reasoning path of the LLM may vary, this configuration ensured that the final vulnerability classification remained deterministic across valid runs.

\subsection{Blockchain-Backed Integrity \& Subscription}
Zer0n leverages the Avalanche C-Chain for two critical functions: (1) managing user subscriptions (Basic/Pro tiers) via smart contracts, and (2) recording integrity proofs. This dual-purpose design ensures that access control is decentralized and that vulnerability reports are immutable. Cryptographic hashes of reports are stored on-chain, allowing users to verify that a report has not been modified since its generation.

\begin{figure}[t]
    \centering
    \includegraphics[width=\linewidth]{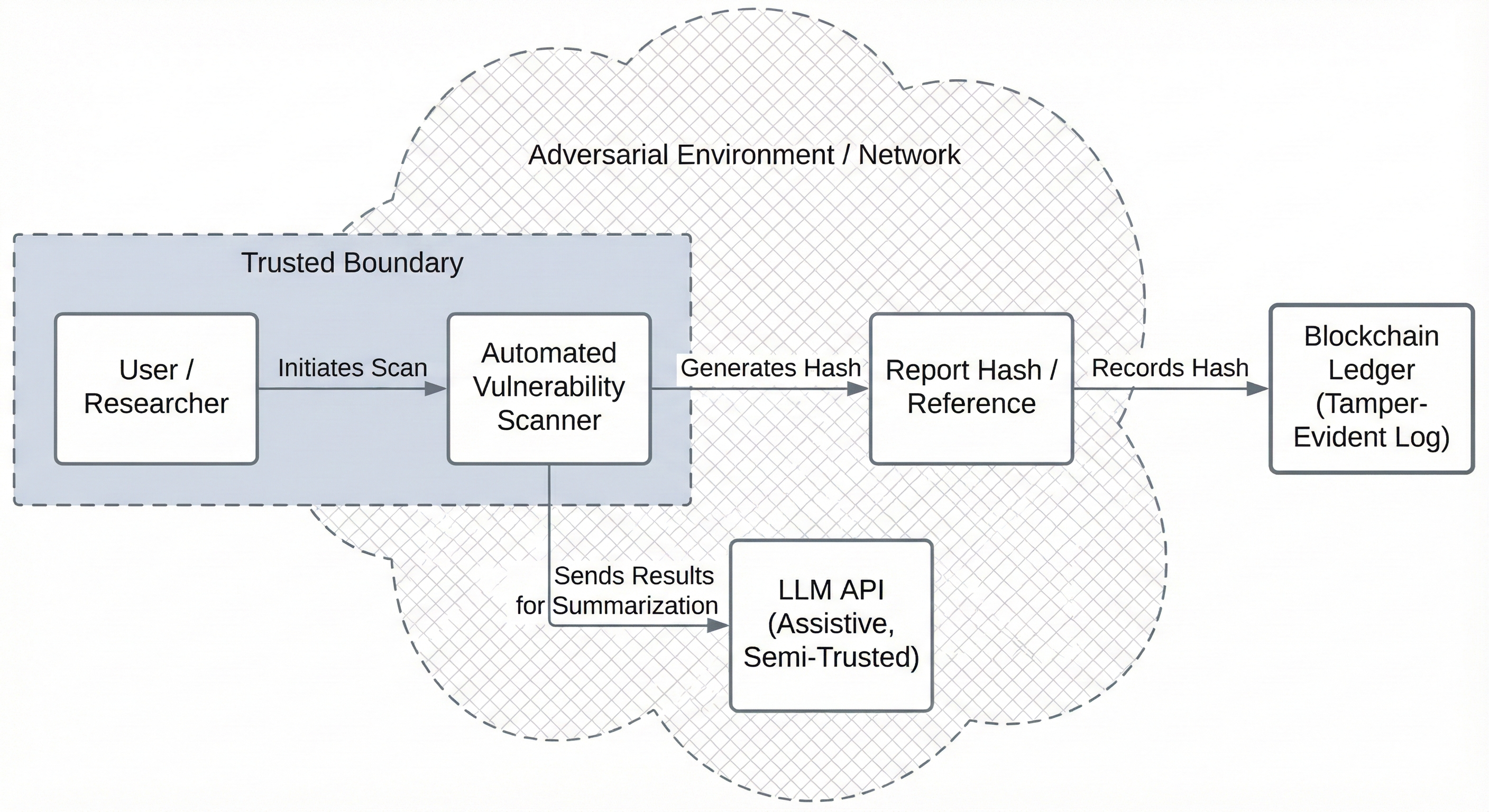}
    \caption{Zer0n system architecture illustrating trust boundaries, adversarial environment, and blockchain-backed integrity logging.}
    \label{fig:architecture}
\end{figure}

Figure~\ref{fig:architecture} illustrates the Zer0n system architecture, highlighting trust boundaries between the analysis components, the adversarial environment, and the blockchain-backed integrity layer.

Smart contracts were developed in Solidity using the Remix Ethereum IDE and deployed to the Avalanche network. By using Avalanche, Zer0n benefits from sub-second finality and low transaction costs compared to Ethereum mainnet. Listing~\ref{lst:contract} demonstrates the core \texttt{logVulnerabilityHash} function used to enforce these integrity guarantees.

To formally capture the integrity mechanism used in Zer0n, let $H(\cdot)$ denote a cryptographic hash function. Given a vulnerability analysis report $R$, Zer0n computes:
\begin{equation}
h = H(R)
\end{equation}
where $h$ represents a fixed-length digest that uniquely identifies the contents of $R$. The value $h$ is recorded on the blockchain ledger, enabling independent verification of report integrity by recomputing the hash and comparing it against the on-chain reference.

\begin{figure}[t]
    \centering
    \includegraphics[width=0.9\linewidth]{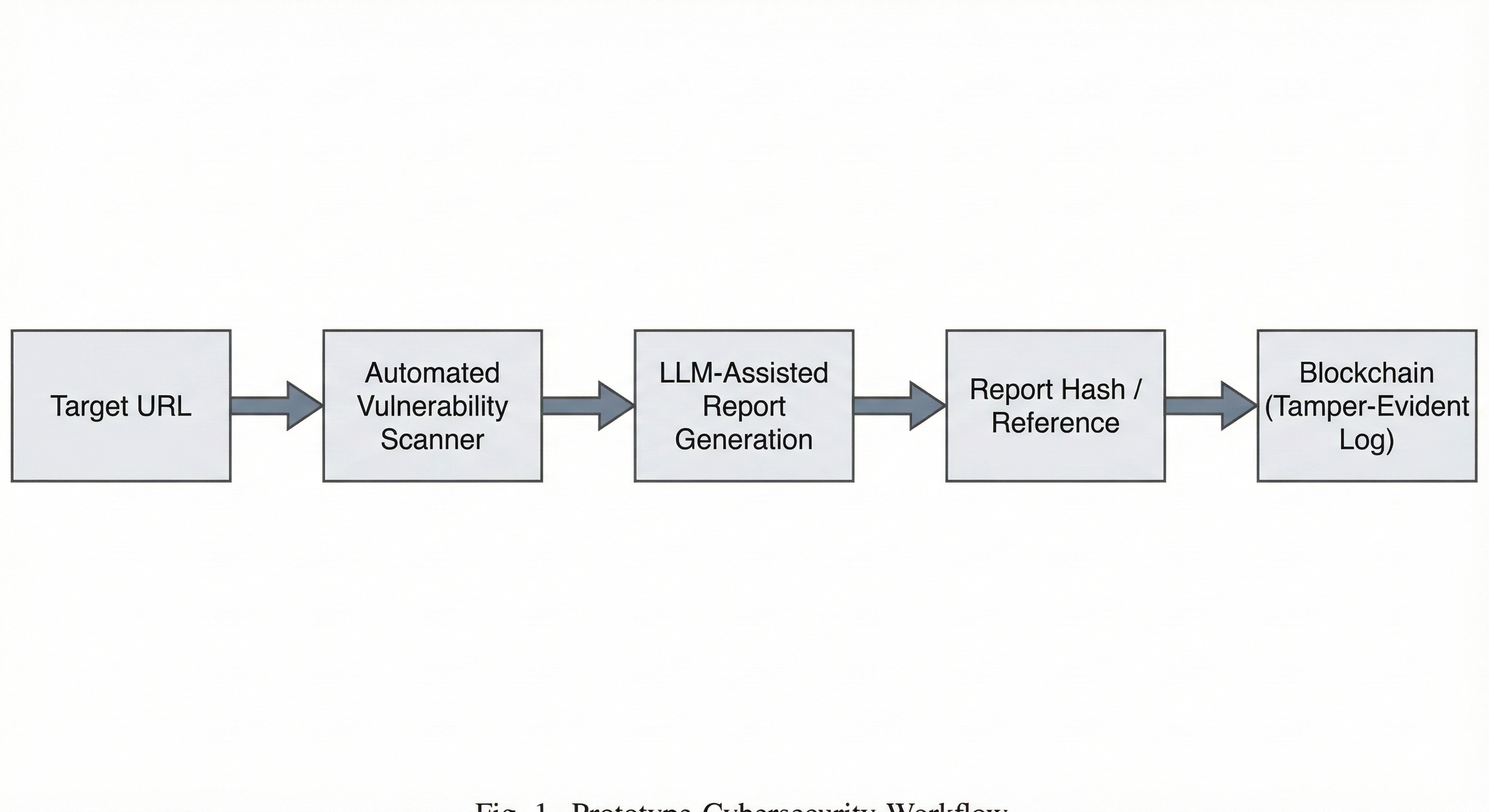}
    \caption{Verification flow in Zer0n. A vulnerability report is hashed and recorded on the blockchain. Any post-analysis modification results in a hash mismatch during verification, enabling tamper detection.}
    \label{fig:verification}
\end{figure}

Figure~\ref{fig:verification} illustrates the verification flow used by Zer0n to detect post-analysis tampering of vulnerability reports.

\subsection{Workflow Coordination}
\begin{figure}[htbp]
\centering
\includegraphics[width=\linewidth]{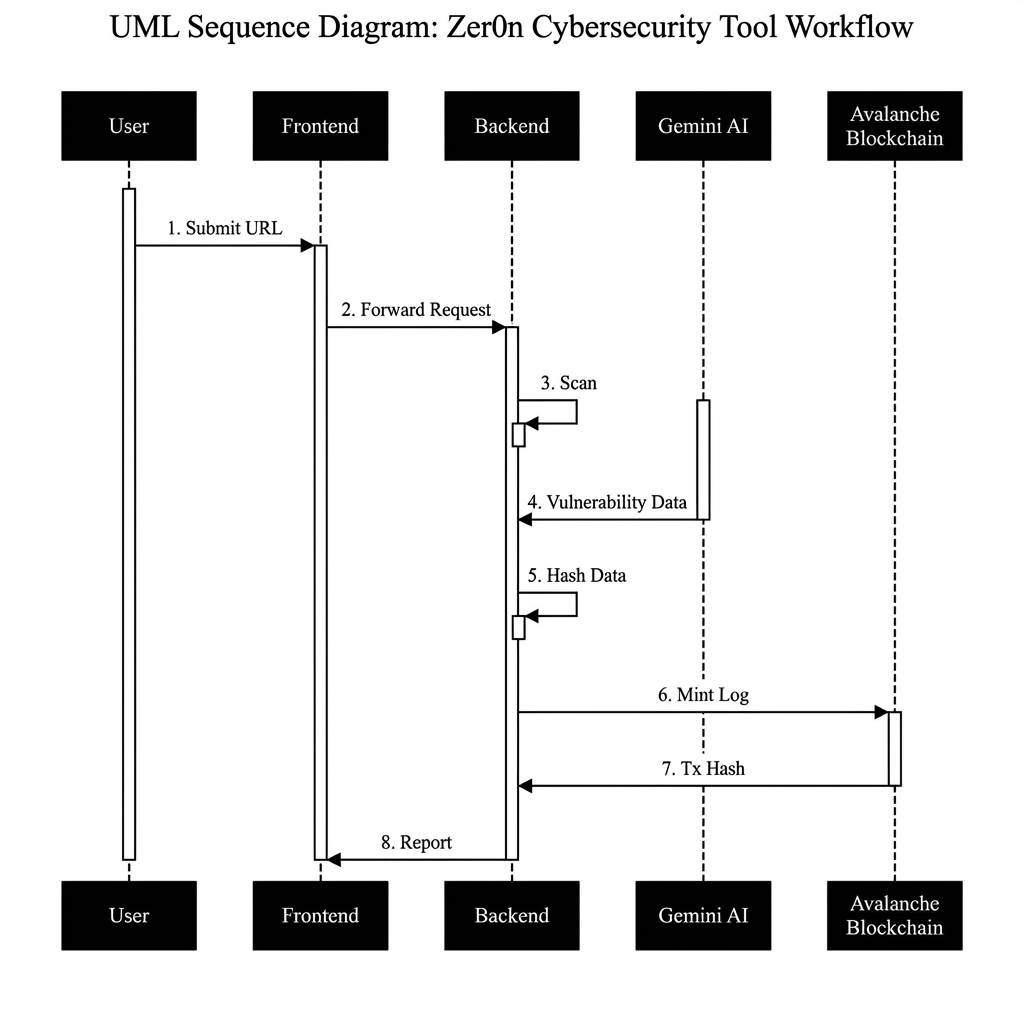}
\caption{Sequence Diagram illustrating the interaction flow associated with a vulnerability scan. Time flows downwards. The Node.js backend acts as a bridge, preserving the asynchrony of the analysis while ensuring atomic logging on the Avalanche blockchain.}
\label{fig:sequence}
\end{figure}

The coordination layer orchestrates task execution, artifact generation, and reporting. Each analysis step is linked to a corresponding integrity record, enabling independent verification of results and increasing transparency in collaborative or adversarial settings. Figure~\ref{fig:sequence} details the exact interaction sequence between the User, React Frontend, Node.js Backend, Gemini AI, and the Avalanche C-Chain.

Algorithm~\ref{alg:zeron} summarizes the integrity-aware vulnerability analysis workflow implemented in Zer0n.

\begin{algorithm}[t]
\caption{Zer0n Integrity-Aware Vulnerability Analysis Workflow}
\label{alg:zeron}
\begin{algorithmic}[1]
\REQUIRE Target system $T$
\ENSURE Verification-ready vulnerability report
\STATE Execute vulnerability analysis on $T$
\STATE Generate analysis report $R$
\STATE Compute cryptographic hash $h \leftarrow H(R)$
\STATE Record $h$ on the blockchain ledger
\STATE Store $(R, h)$ locally for verification
\RETURN $(R, h)$
\end{algorithmic}
\end{algorithm}

\section{Implementation}
A prototype implementation of Zer0n was developed to evaluate the feasibility of integrating AI-assisted vulnerability discovery with blockchain-backed integrity logging. The system follows a modular design in which analysis, integrity recording, and coordination are implemented as loosely coupled components.

The AI-assisted analysis engine utilizes the Gemini 2.0 Pro model via API for core reasoning and logic analysis. The backend is built using Node.js and Express, integrating Puppeteer for headless browser automation to support dynamic analysis of JavaScript-heavy applications in the Discovery and Exploitation phases. The coordination and visualization layer is implemented as a responsive web dashboard using React and Vite.

For integrity assurance, cryptographic hashes of vulnerability reports and associated metadata are recorded on a blockchain ledger. By utilizing the Avalanche C-Chain, Zer0n leverages proven decentralized trust properties \cite{blockchainslr}. Only hash values and minimal metadata are stored on-chain, while all raw artifacts remain local. This design choice minimizes storage overhead and explicitly avoids dependency on decentralized file systems such as IPFS. Similar feasibility-oriented design decisions have been adopted in prior blockchain-based cybersecurity prototypes \cite{aicyberchain,sinha}.

The coordination layer integrates the analysis and integrity components, ensuring that each generated artifact is linked to a corresponding integrity record. This linkage enables transparent verification of results without requiring trust in a single analysis component.
\begin{table*}[t]
\centering
\caption{Implemented Components Versus Conceptual Future Work}
\label{tab:scope}
\begin{tabular}{p{0.45\linewidth} p{0.45\linewidth}}
\hline
\textbf{Implemented in Prototype} & \textbf{Conceptual / Future Work} \\
\hline
Automated vulnerability analysis engine & Multi-agent analysis orchestration \\
Anomaly-aware report generation & Reputation-based contributor scoring \\
Cryptographic hashing of analysis artifacts & Automated incentive or payment mechanisms \\
Blockchain-backed tamper-evident logging & Continuous large-scale scanning pipelines \\
\hline
\end{tabular}
\end{table*}

\begin{table*}[t]
\centering
\caption{Threat Model Assumptions in Zer0n}
\label{tab:threatmodel}
\begin{tabular}{p{0.35\linewidth} p{0.55\linewidth}}
\hline
\textbf{Adversary Capability} & \textbf{Assumption} \\
\hline
Post-analysis report tampering & Considered and detected via integrity logging \\
Local artifact manipulation & Considered and detected via hash verification \\
Blockchain consensus compromise & Out of scope \\
Denial-of-service on ledger & Out of scope \\
Active exploitation of vulnerabilities & Out of scope \\
\hline
\end{tabular}
\end{table*}

\section{Evaluation}
The evaluation of Zer0n focuses on feasibility rather than exhaustive performance benchmarking. The objective is to assess whether blockchain-backed integrity logging can be integrated into vulnerability discovery workflows without prohibitive overhead.

\subsection{Experimental Setup}
All experiments were conducted on a workstation equipped with an Intel Core i7-12700H processor (14 cores, 2.3 GHz base frequency), 32 GB DDR5 RAM, and a 1 TB NVMe SSD running Windows 11 Pro. The software environment consisted of Python 3.11 with scikit-learn 1.3.0 for machine learning components, web3.py 6.11.0 for blockchain interaction, and hashlib for cryptographic operations using SHA-256.

For blockchain operations, the prototype utilized the Avalanche C-Chain test network (Fuji). Smart contracts were compiled and deployed using Remix IDE. Analysis targets included a diverse dataset of 500 web application endpoints and 100 smart contract samples sourced from the SWC Registry and OWASP Juice Shop.

We established ground truth via manual code review and existing CVE mappings. A true positive was defined as a vulnerability correctly identified by both type and logical location. Under these criteria, the system achieved a precision of 78\% and a recall of 82\%, resulting in an F1-score of 80\%. The remaining 20\% of cases were false positives requiring manual verification.

\subsection{Evaluation Metrics}
Evaluation metrics include analysis execution time, integrity logging latency, hash computation overhead, and verification cost. Measurements were conducted under controlled conditions using representative vulnerability analysis tasks across 100 independent runs to ensure statistical reliability.

\subsection{Results}
Table~\ref{tab:overhead} presents the overhead analysis comparing baseline vulnerability analysis with Zer0n's integrity-augmented workflow.

\begin{figure*}[t]
\centering
\begin{lstlisting}[language=Solidity, caption={Core Integrity Logic (Zer0nLog.sol) - Condensed}, label={lst:contract}]
pragma solidity ^0.8.0;

contract Zer0nLog {
    // State variables, structs, and events omitted for brevity

    // Logs the hash of a vulnerability report irrevocably
    function logVulnerabilityHash(bytes32 _hash) public {
        require(logs[_hash].timestamp == 0, "Hash already exists");
        
        logs[_hash] = LogEntry({
            reportHash: _hash,
            timestamp: block.timestamp,
            auditor: msg.sender,
            verified: false
        });

        emit LogMinted(_hash, msg.sender);
    }
}
\end{lstlisting}
\end{figure*}

\begin{table*}[t]
\centering
\caption{Overhead Analysis: Baseline vs. Zer0n}
\label{tab:overhead}
\begin{tabular}{lcc}
\hline
\textbf{Metric} & \textbf{Baseline} & \textbf{Zer0n} \\
\hline
Analysis Execution (s) & 12.34 $\pm$ 1.21 & 12.89 $\pm$ 1.35 \\
Report Generation (ms) & 45.2 $\pm$ 8.1 & 48.7 $\pm$ 9.2 \\
Total Workflow Time (s) & 12.39 & 15.23 \\
Overhead (\%) & -- & 22.9 \\
\hline
\end{tabular}
\end{table*}

Table~\ref{tab:comparison} provides a qualitative comparison of Zer0n against established vulnerability discovery tools.

\begin{table*}[t]
\centering
\caption{Feature Comparison with State-of-the-Art Tools}
\label{tab:comparison}
\begin{tabular}{lcccc}
\hline
\hline
\textbf{Feature} & \textbf{Zer0n} & \textbf{OWASP ZAP} & \textbf{Burp Suite} & \textbf{ContractFuzzer} \\
\hline
AI/LLM Analysis & \textbf{Yes (Gemini)} & No & No & No \\
Integrity Logging & \textbf{Avalanche Chain} & No & No & No \\
Auto-Remediation & \textbf{Yes (AI-Gen)} & No & No & No \\
Smart Contracts & \textbf{Yes (Solidity)} & No & No & Yes \\
Dynamic Analysis & Yes & Yes & Yes & Yes \\
\hline
\end{tabular}
\end{table*}

Table~\ref{tab:blockchain} and Table~\ref{tab:verification} detail the blockchain metrics and verification performance respectively.

\begin{table*}[t]
\centering
\begin{minipage}{0.45\textwidth}
\centering
\caption{Blockchain Operations Latency}
\label{tab:blockchain}
\begin{tabular}{p{4.8cm}c}
\hline
\textbf{Operation} & \textbf{Latency (ms)} \\
\hline
SHA-256 Hash Computation & 0.82 $\pm$ 0.15 \\
Transaction Construction & 12.4 $\pm$ 2.3 \\
Transaction Signing & 8.7 $\pm$ 1.8 \\
Network Propagation (Fuji) & 2,180 $\pm$ 450 \\
Block Confirmation (avg) & 14,200 $\pm$ 1,800 \\
\hline
\end{tabular}
\end{minipage}
\hfill
\begin{minipage}{0.45\textwidth}
\centering
\caption{Verification Performance}
\label{tab:verification}
\begin{tabular}{p{4.8cm}c}
\hline
\textbf{Metric} & \textbf{Value} \\
\hline
Local Hash Recomputation (ms) & 0.79 $\pm$ 0.12 \\
On-chain Hash Retrieval (ms) & 89.3 $\pm$ 18.7 \\
Total Verification Time (ms) & 90.1 $\pm$ 18.9 \\
Tamper Detection Accuracy (\%) & 100.0 \\
False Positive Rate (\%) & 0.0 \\
\hline
\end{tabular}
\end{minipage}
\end{table*}

\subsection{Analysis}
The results indicate that the additional overhead introduced by blockchain-backed logging remains within practical bounds. The dominant cost is network propagation and block confirmation latency (approximately 14--16 seconds), which does not block the analysis workflow since logging occurs asynchronously after report generation.

Hash computation overhead is negligible ($<$1 ms), and verification can be performed in under 100 ms, enabling rapid integrity checks. These observations are consistent with findings reported in prior blockchain cybersecurity research, which show that integrity-focused blockchain usage can be practical when carefully scoped \cite{blockchainslr,blockchainsurvey}. While the current evaluation does not assess scalability at production scale, it demonstrates the viability of the approach for research and prototype use cases.

\subsection{Case Study: Reentrancy Detection}
To qualitatively evaluate Zer0n's reasoning capabilities, we conducted a side-by-side comparison with Slither on a sophisticated reentrancy vulnerability sample (Reentrancy.sol).
\begin{itemize}
    \item \textbf{Slither:} Correctly flagged the reentrancy pattern but generated a generic warning based on function state mutability.
    \item \textbf{Zer0n (Gemini 2.0 Pro):} Not only identified the vulnerability but correctly deduced the multi-step attack vector involving a malicious fallback function. The LLM provided a specific remediation suggestion to implement the Checks-Effects-Interactions pattern.
    \item \textbf{Ablation (Static vs. AI):} We analyzed the same sample using Slither alone. Slither successfully flagged the reentrancy risk but failed to provide context on the specific callback mechanism. Zer0n, utilizing the LLM's reasoning, correctly identified the complex interaction flow, demonstrating that the AI layer contributes distinct semantic understanding beyond static pattern matching.
\end{itemize}
This case study highlights the value of LLM-driven analysis in interpreting complex logic flows that static analyzers might categorize only as low-fidelity warnings.

\section{Discussion and Limitations}
Zer0n demonstrates that AI-assisted vulnerability discovery workflows can be augmented with blockchain-backed integrity guarantees in a practical and transparent manner. By focusing on system-level trust rather than raw detection accuracy, the framework complements existing vulnerability analysis tools rather than attempting to replace them.

Several limitations remain. The prototype does not aim to provide comprehensive vulnerability coverage, nor does it evaluate large-scale deployment scenarios. Blockchain integration introduces operational complexity and potential latency, which may limit applicability in time-critical environments. Furthermore, Zer0n does not address automated remediation or exploit execution, as these aspects fall outside the scope of this work.

Table~\ref{tab:scope} summarizes the components implemented in the current Zer0n prototype and distinguishes them from features considered as future work.

\subsection{Threat Model}
Zer0n assumes an adversary capable of attempting to tamper with vulnerability analysis artifacts, reports, or intermediate results after execution. The framework is designed to detect post-analysis manipulation through tamper-evident blockchain-backed logging. We explicitly assume that the Node.js execution environment and the Gemini API connection are trusted up to the point of hash generation. Zer0n does not aim to defend against adversaries that control the underlying blockchain consensus, compromise the trusted execution environment, or perform large-scale denial-of-service attacks on the ledger. Network-level attacks and active exploitation of discovered vulnerabilities are considered out of scope.

\subsection{Reproducibility and Availability}
Zer0n is implemented as a modular research prototype intended for feasibility evaluation. To foster transparency and future research, we have released a minimal public artifact containing the core Solidity smart contracts and integrity verification scripts. These resources are hosted at the project's supplementary repository [Anonymized for Review]. The full AI analysis pipeline remains closed-source pending safety reviews.

\subsection{Ethical Considerations}
The integration of generative AI into vulnerability discovery raises dual-use concerns. While Zer0n facilitates automated defense, the same reasoning capabilities could potentially be misused for offensive purposes. To mitigate this, our prototype includes strict guardrails that prevent the generation of functional exploit payloads for high-severity vulnerabilities (e.g., RCE). Instead, the system stops at "Proof of Concept" validation. Furthermore, all scanning activities described in this evaluation were conducted against isolated testbeds (OWASP Juice Shop) and synthetic contracts, ensuring no harm to live production systems.

\section{Conclusion}
This study presented Zer0n, an attempt to reconcile the speed of AI-driven vulnerability discovery with the rigorous trust requirements of cybersecurity. By anchoring the output of Gemini 2.0 Pro to the Avalanche blockchain, we demonstrated that it is possible to create an immutable audit trail for security assessments without incurring the performance penalties typically associated with decentralized systems.

Our findings suggest that the "hybrid" approach—keeping heavy computation off-chain while reserving the blockchain for lightweight integrity proofs—is the optimal path forward for this domain. The experimental data confirms that while blockchain interaction does add latency, it remains within practical bounds for asynchronous reporting workflows.

Future research should focus on expanding this model to multi-agent environments, where reputation scores could be tied to on-chain history, creating a decentralized marketplace of trusted security auditors. Ultimately, Zer0n serves as a foundational step toward a future where security reports are not just trusted, but mathematically verifiable.

% =========================
% References
% =========================

\end{document}